\documentclass[12pt]{iopart}

\usepackage{iopams}
\usepackage{graphicx}
\usepackage{bm}
\usepackage{epstopdf}
\usepackage{units}
\usepackage{braket}
\usepackage{pbox}
\usepackage[colorlinks, linkcolor = blue, citecolor = blue, filecolor = black, urlcolor = blue]{hyperref}
\usepackage{soul}

\newcommand{\rmg}{{\rm g}}

\newcommand{\ODB}{{\rm {OD_b}}}
\newcommand{\rb}{r_{\rm b}}
\newcommand{\la}{l_{\rm a}}
\newcommand{\Dp}{\Delta_{\rm p}}
\newcommand{\gr}{\gamma_{\rm r}}
\newcommand{\Dc}{\Delta_{\rm c}}
\newcommand{\Geit}{\Gamma_{\rm EIT}}

\begin{document}

\title{Nonlinear quantum optics mediated by Rydberg interactions}
\author{O. Firstenberg$^1$}
\address{$^1$Department of Physics of Complex Systems, Weizmann Institute of Science, Rehovot 76100, Israel}

\author{C. S. Adams$^2$}
\address{$^2$Joint Quantum Centre (JQC) Durham-Newcastle, Dept. of Physics, Durham University, Durham DH1 3LE, UK.}

\author{S. Hofferberth$^3$}
\address{$^3$5. Physikalisches Institut, Universit\"{a}t Stuttgart, Pfaffenwaldring 57, 70569 Stuttgart, Germany}

\begin{abstract}
By mapping the strong interaction between Rydberg excitations in ultra-cold atomic ensembles onto single photons via electromagnetically induced transparency, it is now possible to realize a nonlinear optical medium which exhibits a strong optical nonlinearity at the level of individual photons. We review the theoretical concepts and the experimental state-of-the-art of this exciting new field, and discuss first applications in the field of all-optical quantum information processing.
\end{abstract}

\maketitle

\section{Introduction}
One remarkable success of advances in ultra-cold Rydberg physics is the realisation of a medium with a large optical nonlinearity at the single photon level \cite{Kuzmich2012b,Vuletic2012,Adams2013}. Highly-excited Rydberg atoms bring something new to the history of optics as they enable quantum nonlinear media where photons are strongly interacting!

This is significant for a number of reasons. For example, previously, it was generally accepted that the prospects for nonlinear all-optical quantum computing were bleak due to the weakness of optical nonlinearities. Consequently the main focus turned towards linear optics quantum computing (LOQC), which exploits measurement to implement gates \cite{Milburn2001}. However, as this is a probabilistic protocol, scaling is a problem. But now Rydberg quantum optics brings the nonlinear approach back into the frame. A fundamental question remains, even if there is a sufficiently large nonlinearity, is this sufficient to build an optical quantum computer \cite{Shapiro2006}? This question can be partially addressed. As we show here, the Rydberg nonlinearity is not only large but different because of the nature of Rydberg blockade \cite{Lukin2001c}. As the interactions between highly-excited Rydberg atoms are long-range, unlike conventional nonlinear optics such as the optical Kerr effect, the Rydberg nonlinearity is also long-range and so standard no-go theorems do not apply. Also interesting on a more fundamental level, is that the realisation of strongly-interacting photons allows us to study exotic quantum many-body states of light such as photon liquids or photon crystals \cite{Vuletic2013b,Fleischhauer2013,Buechler2014,Fleischhauer2015}.

The principle of Rydberg nonlinear optics \cite{Pritchard2013} is simple. The idea is to take the long-range dipolar interaction between highly-excited Rydberg atoms \cite{Saffman2010,Pillet2010} and map it onto a large interaction between photons. The difficulty is to localise a photon to the characteristic length scale of the dipole-dipole interaction, typically a few microns. There is more than one way to achieve this localisation: For a single emitter one can reverse the emission process. However in free space, mode matching between the input field and the dipolar emission pattern is challenging and the efficiency is limited to $\sim 10\%$ \cite{Kurtsiefer2008,Agio2008,Sandoghdar2009,Kurtsiefer2013}. Alternatively, one can use a cavity or waveguide to solve the mode-matching problem. This works well and cavity QED is a well established and extremely successful field where large single photon nonlinearities are possible albeit at the cost of additional complexity of a hybrid system \cite{Kimble2005,Rempe2007,Kimble2008,Vuletic2013}. Third, in an ensemble of atoms the photon localisation or compression occurs naturally (to some extent) due to the phenomenon of slow light \cite{Harris1991,Fleischhauer2005}. A light pulse inside a medium is a mixture of electromagnetic wave and a dipolar excitation, which at the level of single photons we call a polariton \cite{Fleischhauer2000}. The speed of the polariton and the compression ratio are determined by the group index, $n_{\rm g}$, and hence the dispersive response of the medium. To localise a photon, we would like the group index to be as large as possible. The nonlinear response of the Rydberg medium is proportional to the group index and to the strength of the dipole-dipole interactions and, as we show below, both can be extremely large.

\section{A brief history}
The idea of using Rydberg blockade to generate nonclassical states of light appears in the original blockade paper in 2001 \cite{Lukin2001c}. Lukin {\it et al}. write that the ``collective spin states generated by means of dipole blockade $\ldots$, can be transferred from the spin degrees of freedom to the optical field'' allowing the creation of interesting quantum states of light ``without the use of high-$Q$ cavities''. But at the time, the experimental techniques were not sufficiently advanced to make this work, for example, nearly all experiments involving highly-excited Rydberg states used ionisation for detection and no one had observed a coherent atom-light interaction where the presence of the Rydberg state is read-out directly by an optical field. The key turned out to be the technique of electromagnetically induced transparency (EIT), where an additional control field coupling to a third level renders a medium transparent to resonant light. For more details on this phenomenon, which is now routinely exploited in a wide variety of quantum optics experiments, see the excellent review by Fleischhauer {\it et al}. \cite{Fleischhauer2005}.

Most work on EIT has focused on $\Lambda$-type systems, where a strong control laser couples the excitated state to another ground state. But in 2005, Friedler {\it et al}. for the first time discussed the idea that one could instead couple the excited state to a highly-excited Rydberg state which are conveniently metastable. They showed that one could transfer the strong interactions between Rydberg atoms onto the optical transition and thereby realise a photonic phase gate \cite{Kurizki2005}.

The first experiments on EIT to highly-excited Rydberg states with principal quantum numbers up to $n=124$ were reported by Mohapatra {\it et al.}~in 2007 \cite{Adams2007}. Although this was a classical linear optics experiment, the significant result was that the resonances were narrow, and the combined dephasing and decoherence rates did not exceed $\sim 300~$kHz. This was a breakthrough as it showed that potential problems such as ionisation of the Rydberg atoms were not a `show stopper'. The effect of Rydberg blockade on the optical transmission through an ensemble of ultra-cold atoms was first demonstrated in 2010 \cite{Adams2010}. The first experiments demonstrating manipulation of light at the level of single quantum followed in 2012 by Dudin and Kuzmich \cite{Kuzmich2012b}, Peyronel {\rm et al.} \cite{Vuletic2012}, and Maxwell {\rm et al.} \cite{Adams2013}.

In this review we focus on the underlying mechanism of quantum nonlinear optics using interacting Rydberg atoms, on progress since 2012, and on the challenges ahead. But before looking at the quantum nonlinearity, we present a simple classical argument of why Rydberg EIT offers the largest optical nonlinearities ever demonstrated.

\section{Rydberg nonlinear optics}
At the level of a few photons, optical nonlinearities arise when the response of the medium to a second photon is different to the first. This can occur either because the medium cannot absorb or scatter a second photon at the same time --- as in the case of a single emitter --- or because the resonance condition for the second photon is different. This second case is true for both cavity QED and Rydberg ensembles. For Rydberg-mediated nonlinearities, the first photon creates a Rydberg excitation or Rydberg polariton (where the excitation is spread over many atoms), which both result in a shift of the energy of Rydberg states of nearby atoms. If this shift is signifcantly larger than the excitation linewdith, then a second excitation becomes impossible. This process is known as Rydberg blockade \cite{Lukin2001c}. To understand the nonlinear optical response of a Rydberg ensemble it is convenient to start with the case of a single photon or less than one photon such that there are no dipole-dipole interactions, and see how we can map the exaggerated electric field sensitivity of a highly-excited Rydberg atom into a strong optical response. From here it is a small step to imagine that this external field arises due to the proximity of another Rydberg atom and hence another photon.

\subsection{Single-photon Rydberg Kerr effect}
If an optical nonlinearity arises due to a field-dependent shift in the atomic resonance, then to first order the nonlinear response is proportional to the product of the shift and the gradient of frequency dependence of the refractive index, \emph{i.e.}, the dispersion \cite{Boyd1992}. It is convenient to parameterise the gradient in the refractive index in terms of the group refractive index which is defined as
\begin{eqnarray}
n_{\rm g}=1+\omega\frac{\partial n}{\partial \omega}~,
\end{eqnarray}
where $\omega$ is the angular frequency of the light. In a dilute medium where the refractive index is close to unity, we can write $n=1+\textstyle{1\over 2}\chi_{\rm r}$, where $\chi_{\rm r}$ is the real part of the electric susceptibility. Hence for a large group index, $n_{\rm g}=\textstyle{1\over 2}\omega \partial \chi_{\rm r}/\partial \omega$.

Next we consider the shift of the Rydberg level, which for an external electric field has the form of a dc or ac electric Stark shift,
\begin{eqnarray}
\Delta_{\rm Ryd}=-\frac{1}{2}\frac{\alpha {\cal E}^2}{\hbar}~,\end{eqnarray}
where $\alpha$ is the atomic polarizabilty at the frequency of the external field, which can be different to the frequency of the optical field.
This latter point is the key to origin of large nonlinearities in Rydberg ensembles. The polarizability of Rydberg atoms at optical frequencies is small but the polarizability from dc to microwave frequencies can be enormous \cite{Gallagher1994}. This fact has been exploited in microwave cavity QED experiments for decades, where individual Rydberg atoms are used as ultra-sensitive probes able to monitor few-photon intra-cavity fields \cite{Haroche2001}. These low frequency susceptibilities scale as the principal quantum number $n^7$. To qualitatively understand the Rydberg-Rydberg interaction, consider that another Rydberg atom produces a low frequency field ${\cal E}$ proportional to the induced Rydberg dipole which scales as $n^2$. Consequently, when considering only the dipole-dipole term of the interaction Hamiltonian, we obtain a van-der-Waals type interaction scaling as $\alpha{\cal E}^2\sim n^{11}$.

Writing the nonlinear optical response as slope $\partial \chi_{\rm r}/\partial \omega$ times shift $\Delta_{\rm Ryd}$, we get a term that scales quadratically with the field, \emph{i.e.}, a Kerr-like effect,
\begin{eqnarray}
\chi^{(3)}{\cal E}^2&=&\frac{\partial \chi_r}{\partial \omega}\Delta_{\rm Ryd}=-\frac{n_{\rm g}\alpha}{ \omega}{\cal E}^2~.
\end{eqnarray}
So the Kerr nonlinearity $\chi^{(3)}$ is proportional to the product of the group index and the polarizability.
To get a large group index we would like to work close to resonance or even on resonance, but this has the disadvantage that the imaginary part of the susceptibility is also large, giving rise to off-axis scattering and hence loss. The solution is electromagnetically induced transparency (EIT), where an additional control field renders the medium transparent on resonance due to destructive interference between excitation pathways \cite{Fleischhauer2005}. Group indices as large as $10^6$ are possible using EIT in atomic ensembles as first demonstrated in 1999 \cite{Harris1999}. A large group index gives rise to the phenomenon of {\it slow light}, enabling compression of the light pulse inside the medium. In addition, the ability to control the group index enables `storing' and retrieving light pulses, which is the basis of {\it quantum memory} (See \emph{e.g.}, K. Hammerer {\it et al.} for a recent review \cite{Polzik2010}).

By combining EIT and Rydberg states, we get the best of both worlds, \emph{i.e.}, both the largest possible group index and the exaggerated sensitivity of Rydberg state to low frequency fields either applied externally or induced by other nearby Rydberg atoms.

\subsection{Linear EIT susceptibility}
\label{sec:3.2}
\begin{figure}[tb]
\begin{center}
\includegraphics[width=8.6 cm]{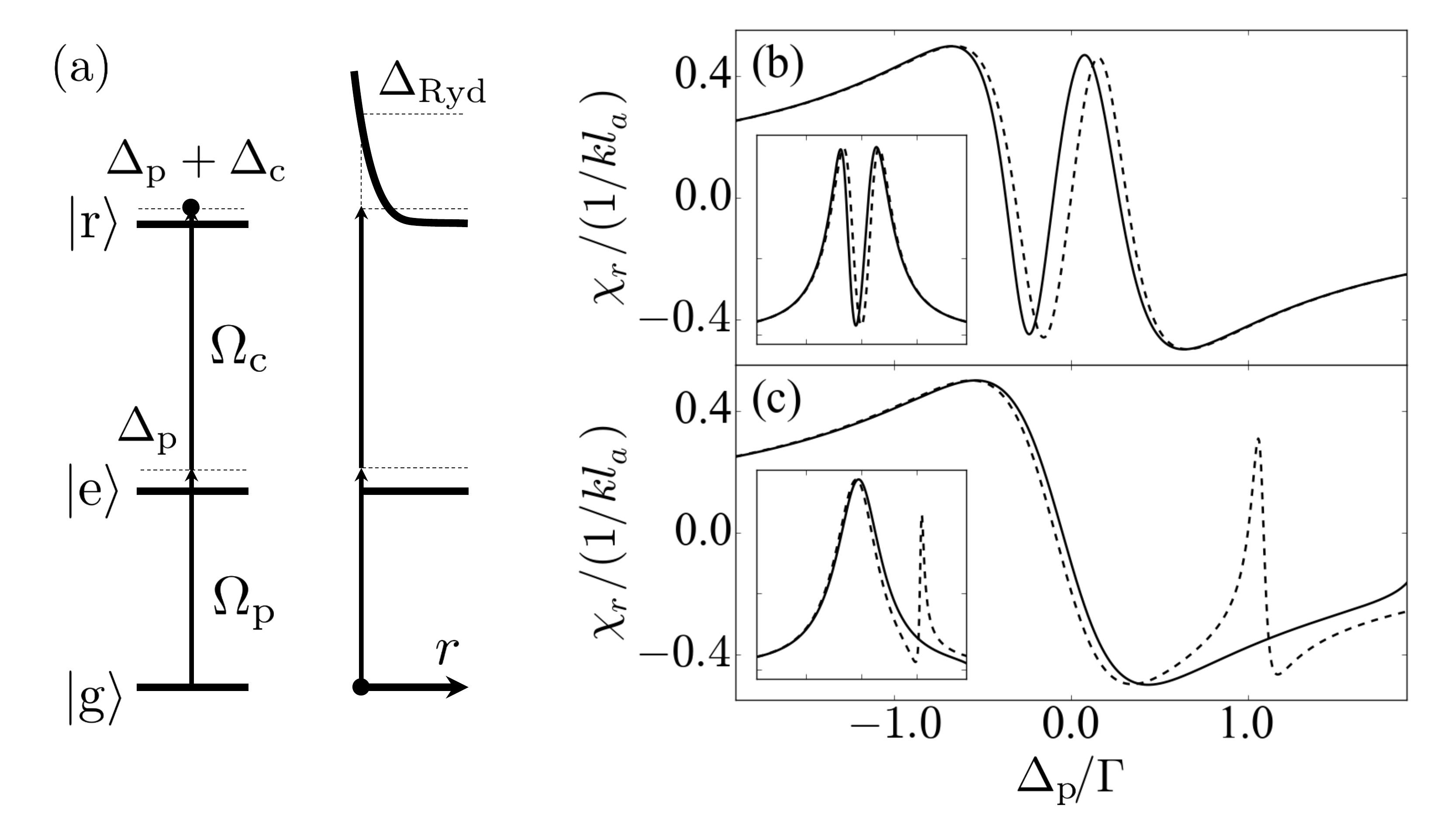}
\end{center}
\caption{Electromagnetically-induced transparency (EIT) with Rydberg states. (a) Level scheme of an unperturbed atom (left) and with a shifted Rydberg level (right). (b+c) The real part of the susceptibility $\chi_{\rm 3-level}$ of an EIT medium from Eq. (\ref{eq:chi_eit}). The insets show the corresponding imaginary part, over the same frequency range. The dashed lines are the unperturbed EIT susceptibilities, with the control field either (b) on resonance or (c) red detuned ($\Dc\approx-\Gamma$) from resonance. The solid lines show the effect of a shift of the Rydberg state: In (b) we demonstrate the effect of a small shift, $\Delta_{\rm Ryd}=-0.1\Gamma$, where the approximation that the nonlinear response equals slope$\times$shift holds. In (c) we demonstrate a large shift, $|\Delta_{\rm Ryd}|>\Gamma$, where the susceptibility reverts to the two-level response, $\chi_{\rm 2-level}$. This corresponds to the case of \emph{Rydberg blockade}. }
\label{fig:chi}
\end{figure}

To see how the large group index arises in an EIT medium, we present a brief derivation of the EIT susceptibility following Gea-Banacloche {\it et al}. \cite{Xiao1995} and Fleischhauer {\it et al.} \cite{Fleischhauer2005}.
Consider the level structure in Fig.~\ref{fig:chi}(a) of an atom with states $\vert {\rm g}\rangle$ and $\vert {\rm e}\rangle$ that is excited by a probe laser with Rabi frequency $\Omega_{\rm p}$ and detuning $\Delta_{\rm p}$. The excited state $\vert {\rm e}\rangle$ is coupled to a highly-excited Rydberg state $\vert {\rm r}\rangle$  by a coupling laser with detuning $\Delta_{\rm c}$ and Rabi frequency $\Omega_{\rm c}$. The equations for the coherences of this 3-level ladder system are
\begin{eqnarray}
\dot{\tilde{\rho}}_{\rm eg}
&=&-{\rm i}\frac{\Omega_{\rm p}}{2}(\tilde{\rho}_{\rm gg}-\tilde{\rho}_{\rm ee})
+{\rm i}\Delta_{\rm p}\tilde{\rho}_{\rm eg}
-{\rm i}\frac{\Omega_{\rm c}}{2}\tilde{\rho}_{\rm rg}-\gamma\tilde{\rho}_{\rm eg}~\nonumber\\
\dot{\tilde{\rho}}_{\rm rg}
&=&{\rm i}\frac{\Omega_{\rm p}}{2}\tilde{\rho}_{\rm re}
+{\rm i}(\Delta_{\rm p}+\Delta_{\rm c})\tilde{\rho}_{\rm rg}
-{\rm i}\frac{\Omega_{\rm c}}{2}\tilde{\rho}_{\rm eg}
-\gr\tilde{\rho}_{\rm rg}~.
\nonumber\\
\dot{\tilde{\rho}}_{\rm re}
&=&-{\rm i}\frac{\Omega_{\rm c}}{2}(\tilde{\rho}_{\rm ee}-\tilde{\rho}_{\rm rr})
+{\rm i}\Delta_{\rm c}\tilde{\rho}_{\rm re}
+{\rm i}\frac{\Omega_{\rm p}}{2}\tilde{\rho}_{\rm rg}
-\gamma'\tilde{\rho}_{\rm re}~,\nonumber
\end{eqnarray}
where $\gamma$ and $\gamma'$ are decoherence rates of the driven transitions ($g\leftrightarrow e$ and $e\leftrightarrow r$), which are often much larger than the decoherence rate $\gr$ of the two-photon transition $g\leftrightarrow r$. If spontaneous emission is the only decay mechanism then $\gamma=\Gamma/2$, where $\Gamma$ is the spontaneous decay rate of state $\vert {\rm e}\rangle$. For the steady-state solution in the weak-probe limit (where the populations $\rho_{\rm gg}=1$, $\rho_{\rm ee}=0$, $\rho_{\rm rr}=0$), we find that $\tilde{\rho}_{\rm re}=0$ and then from the second equation
\begin{eqnarray}
\tilde{\rho}_{\rm rg} = -\frac{{\rm i}\Omega_{\rm c}/2}{\gr-{\rm i}(\Delta_{\rm p}+\Delta_{\rm c})}\tilde{\rho}_{\rm eg}~,\nonumber
\end{eqnarray}
and if we substitute this into the steady-state solution of the first equation, we find an expression for the coherence on the probe transition
\begin{eqnarray}
\tilde{\rho}_{\rm eg} = -\frac{{\rm i}\Omega_{\rm p}/2}{\gamma-{\rm i}\Delta_{\rm p}+\Omega_{\rm c}^2/[\gr-{\rm i}(\Delta_{\rm p}+\Delta_{\rm c})]/4}~,\nonumber
\end{eqnarray}
which determines the induced dipole %$d_{\rm ab}=\alpha_{\rm p}{\cal E}_{\rm p}$
on the probe transition. The resulting electrical susceptibility is
\begin{eqnarray}
\chi_{\rm 3-level}=\frac{2\gamma}{k \la}\frac{\tilde{\rho}_{\rm eg}}{\Omega_{\rm p}}=\chi_{\rm 2-level}\left[1-\frac{\Omega_{\rm c}^2}{4(\gr-{\rm i}\Dp-{\rm i}\Dc)(\gamma-{\rm i}\Dp)+\Omega_{\rm c}^2}\right]~,
\label{eq:chi_eit}
\end{eqnarray}
with the susceptibility of the bare two-level system ($\Omega_{\rm c}=0$) given by
\begin{eqnarray}
\chi_{\rm 2-level}=\frac{1}{k \la} \frac{{\rm i}\gamma}{\gamma-{\rm i}\Delta_{\rm p}}~.
\label{eq:chi_2}
\end{eqnarray}
Here $\la= (N\sigma)^{-1}$ is the resonant attenuation length, with $N$ the number density of atoms and $\sigma$ the optical cross-section ($\sigma=3\lambda^2/2\pi$ for a closed two-level transition). These susceptibilities are plotted in Figs.~\ref{fig:chi}(b,c).

To simplify the presentation, we now focus on the resonance case $\Dc=0$. Whereas in the two-level system we have maximum scattering on resonance, in EIT the scattering is suppressed. Minimum scattering is obtained at two-photon resonance $\Dp=-\Dc=0$, where
\begin{equation}
\chi_{\rm 3-level}(\Dp=-\Dc=0)=\chi_{\rm 2-level}\left[1-\frac{\Omega_{\rm c}^2}{4\gr\gamma+\Omega_{\rm c}^2}\right]~,
\end{equation}
so we require $\Omega_{\rm c}^2 \gg 4\gr\gamma$ to induce significant transparency. In this regime, the EIT linewidth $\Geit=\Omega_{\rm c}^2/(4\gamma)$ is dominated by power broadening ($\Geit\gg\gr$). Assuming an EIT transparency window much narrower than the one-photon absorption line ($\Geit\ll\gamma$), we find from Eq.~(\ref{eq:chi_eit})
\begin{equation}
\chi_{\rm 3-level}\approx\chi_{\rm 2-level}\left(1-\frac{\Geit}{\Geit-{\rm i}\Dp}\right)~
\end{equation}
for probe tuned within the EIT linewidth ($|\Dp|\ll\Geit$ and $\Dc=0$).
%with the EIT linewidth given by $\Geit=\Omega_{\rm c}^2/(4\gamma)$.
If we now linearize the susceptibility around $\Delta_{\rm p}=0$, we find a real part $\chi_{\rm r}\approx(k \la)^{-1} \Dp / \Geit$
%\st{which is the same as the 2--level case with the EIT linewidth replacing the 2--level linewidth}
and a corresponding group index
\begin{eqnarray}
n_\rmg\approx\frac{\omega}{2}\frac{\partial \chi_{\rm r}}{\partial \Dp} &\approx& \frac{1}{k \la}
\frac{\omega}{2\Geit}~.%{\color{yellow}\approx \frac{c}{\la}\frac{2\gamma}{\Omega_{\rm c}^2}}
\label{eq:n_g_EIT}
\end{eqnarray}
%\st{Thus EIT enables a larger group index and hence larger optical nonlinearity when the EIT linewidth is less than the 2--level linewidth, as shown in Fig. 1. This is satisfied by our assumptions above, which require that $\gamma\gg\gr$, hence state $\vert {\rm r}\rangle$ must be much longer lived than state $\vert {\rm e}\rangle$ which is the case if $\vert {\rm r}\rangle$ is another ground state ($\Lambda$--EIT) or a Rydberg state.}
Thus narrow EIT resonances enable a large group index and hence large optical nonlinearity. To obtain a narrow resonance (small $\Geit$) while satisfying the above requirement $\Geit\gg\gr$ to guarantee significant transparency, we require a long-lived state $\vert {\rm r}\rangle$ (small $\gr$). This is indeed the case if $\vert {\rm r}\rangle$ is another ground state ($\Lambda$--EIT) or a Rydberg state.

The nonlinear response arises from a level shift which changes $\Delta_{\rm c}$, as illustrated in Fig.~\ref{fig:chi}. The success of Rydberg nonlinear optics relies on the ability to map the large shifts arising from low frequency fields onto an optical field using EIT. The first experiment demonstrating a large Kerr effect due to an external electric field using Rydberg EIT was reported in 2008 \cite{Adams2008}.

\subsection{Optical nonlinearity due to Rydberg blockade}
The probe field propagates in the Rydberg-EIT medium as a so-called Rydberg polariton with a group velocity
\begin{equation}
v_\rmg=\frac{c}{n_\rmg}.
\end{equation}
A large group index thus implies a small photonic component, on order $v_\rmg/c$, and correspondingly a large Rydberg component. The interaction between the Rydberg atoms shifts the Rydberg level, effectively altering the control-field detuning $\Dc$. The term \emph{Rydberg blockade} refers to the case where the interaction-induced shift is much larger than the EIT linewidth. In this case, the nonlinearity can be considered as a switch from the 3-level EIT susceptibility $\chi_{\rm 3-level}$ to the 2-level susceptibility $\chi_{\rm 2-level}$ \cite{Pohl2011b,Lukin2011,Fleischhauer2011}, as illustrated in Fig.~\ref{fig:chi}(c).

The volume around a Rydberg polariton in which EIT is suppressed is known as the blockade sphere. Its radius is found from the requirement $V(\rb)=2\hbar\Geit$, where $V(r)$ is the Rydberg-Rydberg interaction potential. For a van der Waals interaction $V(r)=C_6/r^6$, we find
\begin{equation}
\rb=\sqrt[6]{C_6/(2\hbar\Geit)}.
\end{equation}
If there are enough atoms contributing to the 2-level susceptibility within the blockade sphere, the effect of a single Rydberg polariton on the transmission of nearby photons can be dramatic. This is the quantum nonlinear regime that will be the focus of the rest of this article and requires that the medium has a high optical depth per blockade sphere.

However, even in the `classical' -- or partially blockaded -- regime, the non-linearities can be enormous! For a weak probe, $\Omega_{\rm p}\ll \Omega_{\rm c}$, as the probe intensity increases there is a gradual transition from the 3--level to the 2--level response as each successive Rydberg excitation converts a fraction $(\Omega_{\rm p}/\Omega_{\rm c})^2$ of nearby atoms to 2--level scatterers. It follows that the classical Rydberg non-linearity scales as the 2--level response times the fraction of blockading excitations  \cite{Pohl2011b}:
\begin{eqnarray}
\chi^{(3)}&=&N\frac{4\pi}{3}r_{\rm b}^3\frac{\Omega_{\rm p}^2}{\Omega_{\rm c}^2}\chi_{\rm 2-level}~.
\end{eqnarray}
Substituting $\Omega_{\rm p}=d{\cal E}_{\rm p}/\hbar$, where $d$ is the dipole matrix element for the 2--level transition, this equation gives a {\it Kerr-like optical non-linearity}.  For $\rb=5~\mu$m (typical for principal quantum numbers $50-100$ and $\Omega_c$ on order of a few MHz, satisfying $\Omega_{\rm c}^2 \gg 4\gr\gamma$) and $N=3\times 10^{12}~{\rm cm}^{-3}$ in Rb, one obtains an estimate of the Rydberg nonlinearity of $\chi_{\rm 3-level}\sim 5\times 10^{-2}~{\rm V}^{-2}{\rm m}^2$, which is 5 orders of magnitude larger than conventional EIT media \cite{Fleischhauer2005}. This classical nonlinearity was first demonstrated experimentally in 2010 using rubidium atoms prepared at densities of
$N=2\times 10^{10}~{\rm cm}^{-3}$ using a magneto-optical trap \cite{Adams2010}.

Having reviewed the classical nonlinearity arising from Rydberg EIT we now move on to the quantum limit.

\section{Quantum nonlinearity}
Quantum nonlinear optics \cite{Lukin2014b}, that is, the extreme limit where the optical nonlinearity becomes significant on the level of single photons, calls for a quantum description of the probe field. In this limit, the nonlinearity is characterized by comparing the transmission amplitude of single photons to that of a photon pair. If single photons are transmitted much better than photon pairs, or conversely absorbed (scattered) much more, we denote the nonlinearity as \emph{dissipative}. This type of nonlinearity can be quantified, almost by definition, by measuring the normalized second-order correlation function $g^{(2)}(t_1,t_2)=\langle n(t_1)n(t_2) \rangle / [ \langle n(t_1)\rangle \langle n(t_2) \rangle ]$ of the outgoing field for a weak classical input. Here, $t_1$ and $t_2$ are the photon detection times, and $n(t)$ the detection rate. For example, anti-bunching $g^{(2)}(t_1=t_2) \ll 1$ would indicate strong scattering of pairs (the numerator of $g^{(2)}$) relatively to singles (approximately the denominator). The other type of nonlinearity is the so-called \emph{dispersive}, when singles and pairs are equally transmitted (\emph{e.g.}, in a lossless medium) but acquire different optical phases. Strong nonlinearity is then reached when the phase difference, sometimes refereed to as the \emph{conditional} or \emph{nonlinear phase} $\phi$, is on the order of $\pi$. We discuss below how one could generalize the standard $g^{(2)}$ measurement to characterize the conditional phase.

In a Rydberg-EIT setup, the type of the nonlinearity is determined by the detunings $\Dp$ of the probe from the intermediate level $\vert e\rangle$ \cite{Lukin2011}. Each polariton, carrying a single Rydberg excitation, suppresses EIT within the blockade sphere around it. The transmission amplitude for other photons crossing this blockade sphere is thus given by the bare two-level response
\begin{equation}
t_{\rm 2-level}(\Dp)=\exp\left (- \frac{\gamma}{\gamma-{\rm i}\Dp} \frac{\ODB}{2} \right). \label{eq:t2}
\end{equation}
Here the optical depth of the blockade sphere $\ODB=2\rb/\la$ depends on the sphere diameter $2\rb$ and the attenuation length $\la$. The quantum nonlinearity follows from the ratio $t_{\rm 2-level}/t_{\rm 3-level}$, which can be estimated from the classical 2- and 3-level susceptibilities of the form plotted in Figs.~\ref{fig:chi}(b,c).Since $t_{\rm 3-level}=1$ at the EIT resonance (assuming $\Geit \gg \gr$), we only need to examine $t_{\rm 2-level}$: For a resonant probe, $t_{\rm 2-level}(\Dp=0)=\exp(-\ODB/2)$ leads to scattering of blocked photons and thus to dissipative nonlinearity. For an off-resonant probe, $t_{\rm 2-level}(|\Dp|\gg \gamma)\approx\exp({\rm i}\phi)$ yields no loss and a conditional phase of $\phi=-(\ODB/2)(\gamma/\Dp)$, thus rendering a dispersive nonlinearity.

\subsection{Dissipative nonlinearities}
\label{subsec:dissipative_nonlinearities}
It follows from the above expressions that the strength of the nonlinearity is governed by $\ODB$, and we require $\ODB\ge1$ for the \emph{quantum} nonlinear limit, in both the dissipative and the dispersive regimes. A medium with $\ODB\approx 5$ was realized in 2012 by Peyronel {\it et al.}~\cite{Vuletic2012}. In this experiment, an elongated optical trap compressed a cloud of ultracold rubidium to an atomic density of $N= 2\times 10^{12}~{\rm cm}^{-3}$. The Rydberg-EIT had a (half) linewidth of $\Geit/(2\pi)\approx 10~{\rm MHz}$ when tuned to the Rydberg level 100S$_{1/2}$, yielding a blockade radius on order $\rb \approx 10~\mu{\rm m}$ and approximately 10,000 atoms within a single blockade sphere. The probe field was focused throughout the $100~\mu{\rm m}$-long cloud to a diameter $2w_0$ on the order of $\rb$. By that, the condition $w_0\ll\rb$, for keeping the dynamics one-dimensional and blocking two polaritons from propagating side-by-side, was nearly fulfilled.

\begin{figure}[tb]
\begin{center}
\includegraphics[width=8.6 cm]{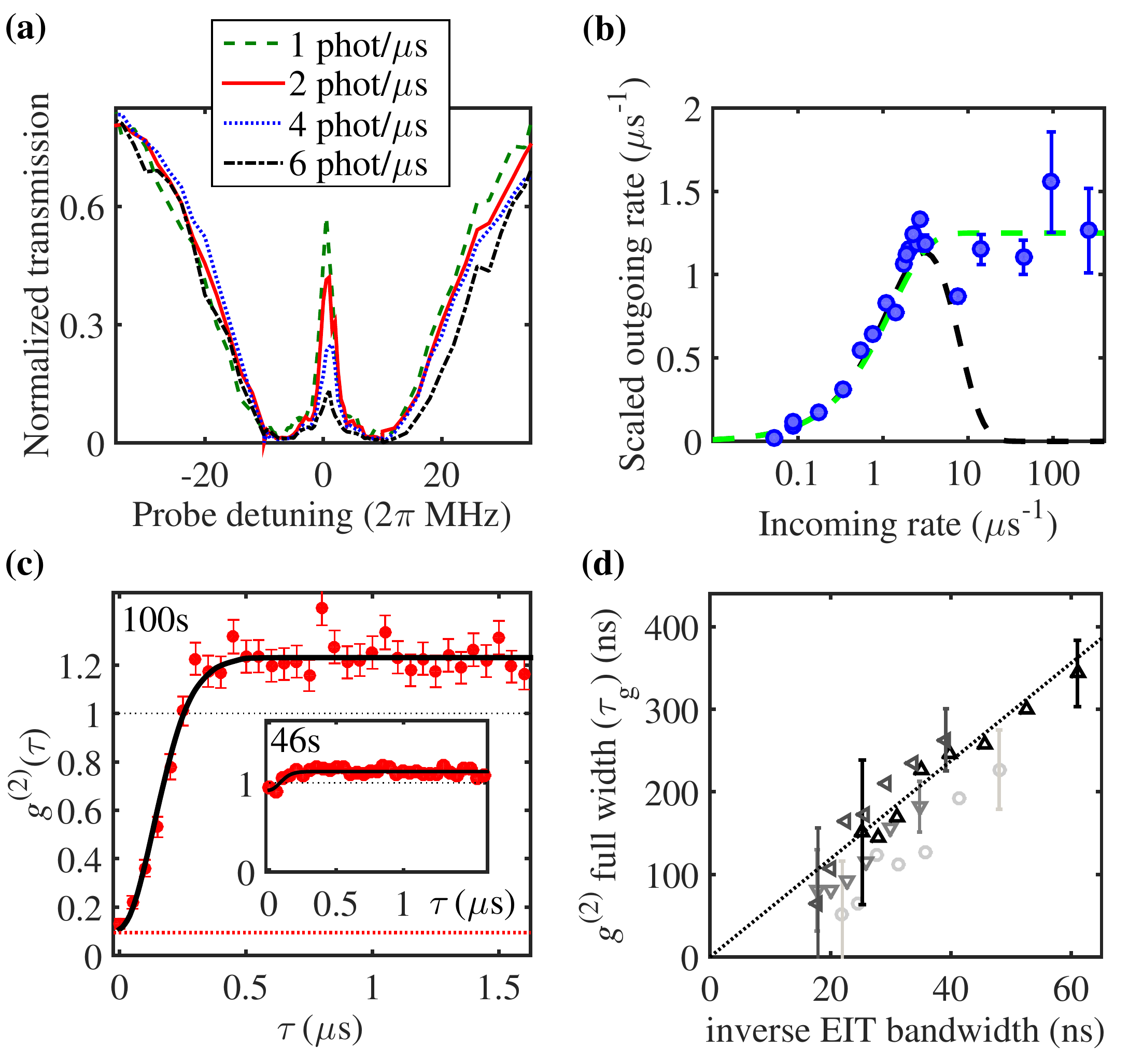}
\end{center}
\caption{(a) Transmission spectra of EIT using the Rydberg state 100S$_{1/2}$ for various incoming photon rates. The weak field transmission is determined by the 3-level susceptibility, Eq.~(\ref{eq:chi_eit}), plotted in Fig.~\ref{fig:chi}(b) inset. However, the transmission on resonance begins to be substantially reduced at a level of a few photons per $\mu$s. With the group delay in the medium being $\tau_{\rm d}\approx 0.25~\mu$s, this rate corresponds to (on average) less than two photons inside the medium.
(b) Outgoing photon rate (rescaled to compensate for the $50\%$ linear transmission and for the finite detection efficiency) versus incoming photon rate. The transmission is saturated at about one photon per $\mu$s. The dashed curves are expected rates assuming that multi-photon events are either (black) blocked or (green) converted into a one-photon state.
(c) Normalized second-order correlation function $g^{(2)}$ of the outgoing photons versus their time separation $\tau$. The anti-bunching feature, a result of the dissipative nonlinearity, has a temporal width $\tau_{\rm c}$ on the order of the group delay $\tau_{\rm d}$. Inset: a reference experiment with the Rydberg state 46S$_{1/2}$, showing a negligible effect. The solid lines are results of full numerical simulations of the 2-photon wavefunction evolution in the medium. (d) The anti-bunching temporal width $\tau_{\rm c}$ versus the inverse bandwidth $B=\Geit/\sqrt{8{\rm OD}}$ for various experimental parameters (different symbols). The finite transmission bandwidth $B$ broadens the anti-bunching feature in the two-photon wavefunction during propagation. Figure adapted from Peyronel {\it et al.}~\cite{Vuletic2012}.}
\label{fig:dissipativeQNLO}
\end{figure}

In these conditions, nonlinear transmission was measured at probe powers as low as 0.25~pW, corresponding to less than two photons in the medium --- see Fig.~\ref{fig:dissipativeQNLO}(a) and caption. For incoming photon rates above $2~\mu{\rm s}^{-1}$, the outgoing photon rate became constant [Fig.~\ref{fig:dissipativeQNLO}(b)], realizing a photonic version of an hourglass which allows the transmission of only about one photon per $\mu$s. The quantum nature of the nonlinearity is evidenced by a strong photon anti-bunching of the outgoing light [Fig.~\ref{fig:dissipativeQNLO}(c)], measured with single-photon detectors. In the regime of the experiment, the (linear) transmission bandwidth of the medium, which sets a lower limit on the duration of probe pulses, sets a similar lower limit on the temporal extent of the anti-bunching feature [Fig.~\ref{fig:dissipativeQNLO}(d)].

\subsection{Dispersive nonlinearities}
\label{subsec:dispersive_nonlinearities}
\begin{figure}[tb]
\begin{center}
\includegraphics[trim={0cm 0.7cm 0cm 1.0cm},clip,width=8 cm]{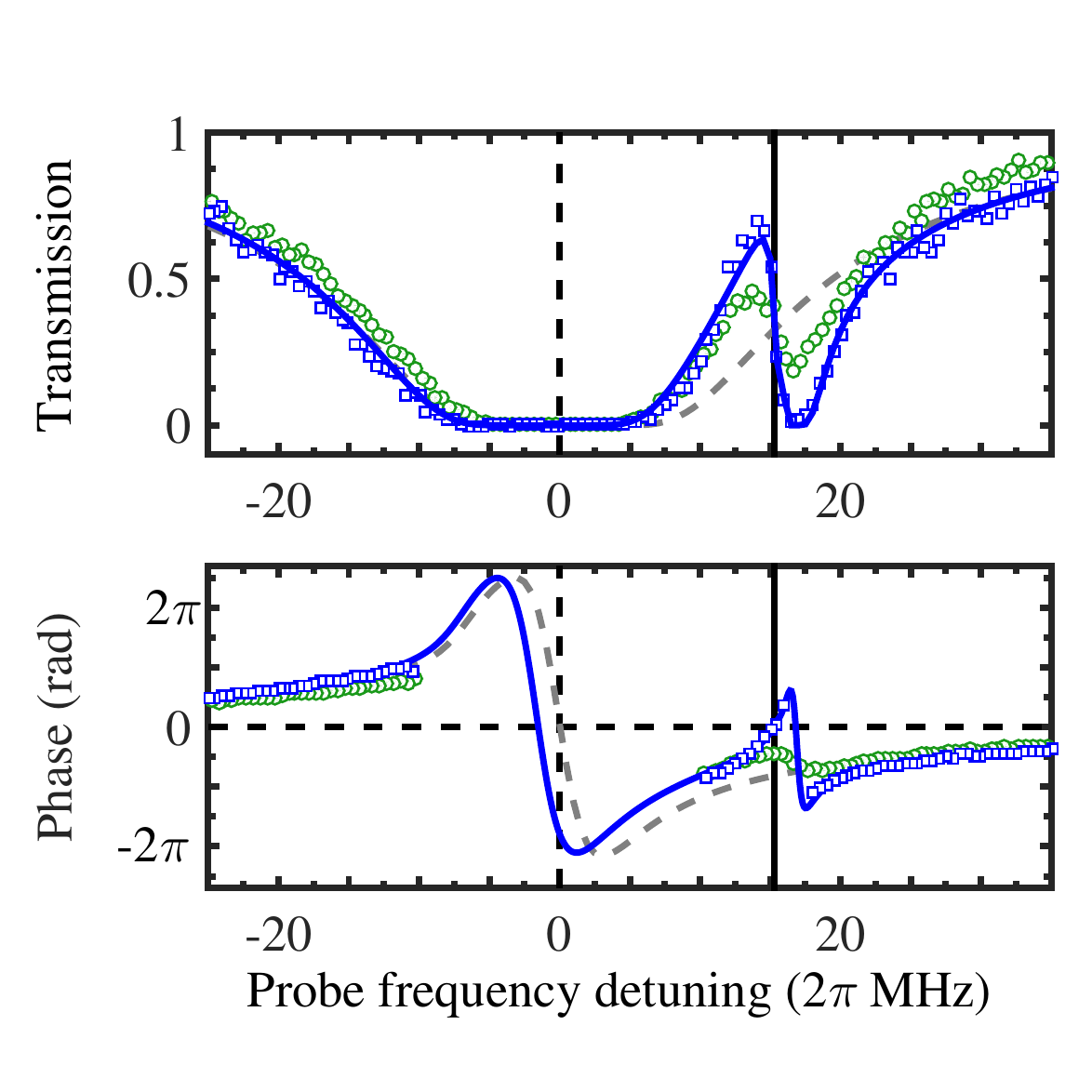}
\end{center}
\caption{Transmission (top) and phase shift (bottom) versus the probe detuning $\Dp$ when the control is detuned by $15$ MHz from resonance. An incoming rate of 0.5 photons per $\mu$s (blue squares) is compared to a higher rate of 5 per $\mu$s (green circles). Dispersive nonlinearity is obtained at a probe detuning corresponding to the black vertical line: the transmission depends only weakly on the incoming rate, while the phase shift exhibits a substantial change. The theoretical lines correspond to the (solid blue) 3-level susceptibility $\chi=\chi_{\rm 3-level}$ and (dashed gray) 2-level susceptibility $\chi=\chi_{\rm 2-level}$, given in Eqs.~(\ref{eq:chi_eit}) and (\ref{eq:chi_2}) and similar to the case plotted in Fig.~\ref{fig:chi}(c); Here, the transmission is given by $\exp[-k\la{\rm Im}(\chi){\rm OD}]$ and the phase by $k\la{\rm Re}(\chi){\rm OD}/2$. Evidently, the measured response approaches that of the 2-level system for higher incoming photon rates. Figure adapted from Firstenberg {\it et al.}~\cite{Vuletic2013b}.}
\label{fig:dispersiveQNLO1}
\end{figure}

Changing the nonlinearity in a Rydberg-EIT experiment from dissipative to dispersive is straight forward: one simply detunes the probe and control fields from the intermediate state. Firstenberg {\it et al.}~realized this in 2013 with detunings $\Dp\approx-\Dc\approx 5\gamma$ \cite{Vuletic2013b}. The measured transmission and phase-shift spectra, corresponding respectively to the imaginary and real parts of the susceptibility, are shown in Fig.~\ref{fig:dispersiveQNLO1}. At the chosen detuning, the 3-level and 2-level responses differ only in their phase, yielding purely dispersive nonlinearity.

\begin{figure}[tb]
\begin{center}
\includegraphics[trim={0cm 3.5cm 0cm 3.5cm},clip,width=8.4 cm]{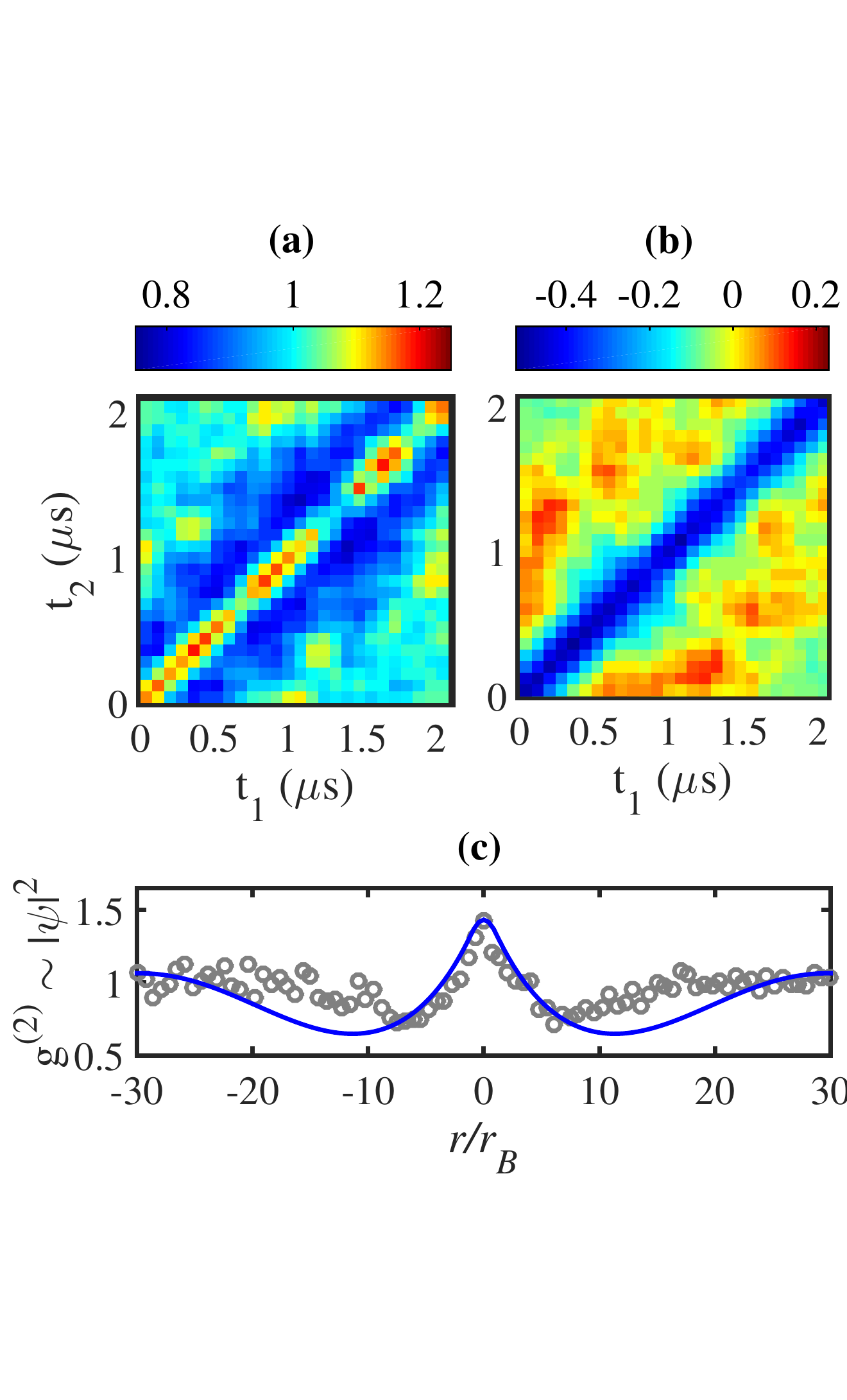}
\end{center}
\caption{Dispersive quantum nonlinearity at $\Dp=4.6\gamma$. (a) Normalized second-order correlation function $g^{(2)}(t_1,t_2)$ for photons detected at times $t_1$ and $t_2$, indicating the bunching of photons. (b) Conditional phase-shift $\phi(t_1,t_2)$ (color scale in radians), extracted from correlation measurements with reference non-interacting photons. (c) Measured $g^{(2)}(\tau=t_1-t_2)$ (gray circles), where time has been converted to distance via $r=v_\rmg\tau$, compared to the solution of the Schr\"odinger equation (\ref{eq:RydSch}) (blue line). The bunching observed in (a) and (c) is governed by the 2-photon bound state of this approximated Schr\"odinger evolution. Figure adapted from Firstenberg {\it et al.}~\cite{Vuletic2013b}.}
\label{fig:dispersiveQNLO2}
\end{figure}

Similarly to the case of dissipative nonlinearity, the measured spectra alone do not suffice to characterize the quantum nonlinearity, and 2-photon correlation measurements are needed. Since now it is a phase, rather than transmission, that is of interest, an interferometric setup is required. To this end, one introduces a reference photonic mode, \emph{e.g.,} with a different polarization or frequency, for which the transmission is linear. By measuring the correlations between the probe and the reference photons when interfered in different bases, one can utilize quantum tomography techniques to reconstruct the full two-photon density matrix $\rho(t_1,t_2)$ (with $t_1$ and $t_2$ being the detection times of the two photons). Normalizing this density matrix by the one-photon density matrices $\rho(t_1)\otimes\rho(t_2)$ renders an \emph{interaction matrix} $\tilde{\rho}(t_1,t_2)$ that generalizes the standard $g^{(2)}(t_1,t_2)$. The interaction matrix, $\tilde{\rho}(t_1,t_2)$, yields not only the conditional phase $\phi$, but also information on the decoherence and entanglement generation during the process. Experimental results extracted from $\tilde{\rho}(t_1,t_2)$ in the 2013 experiment \cite{Vuletic2013b} are shown in Figs.~\ref{fig:dispersiveQNLO2}(a-b). A conditional phase shift as high as $|\phi|=\pi/4$ was observed in this experiment at a linear transmission of $50\%$.

\subsection{Photon-photon interaction}
The optical nonlinearity we observe originates from the strong dipolar interaction between Rydberg atoms. A reciprocal and complimentary notion is the \emph{effective interaction between photons} that can be used to describe the nonlinearity at the quantum level. In fact, this effective photon-photon interaction inherits from the dipolar atomic interaction, but is regulated by the optical response of the medium. Dissipative and dispersive nonlinearities thus result from effective dissipative or dispersive photon-photon interactions.

To illustrate this point, we define the 2-photon wavefunction $\psi(z_1,z_2)$ \cite{Vuletic2013b}, with $z_1$ and $z_2$ the photon coordinates inside the medium. $|\psi|^2$ and ${\rm arg}(\psi)$ are respectively the probability to find the two photons and their phase-shift relatively to the non-interacting (Poissonian) case. In the absence of nonlinearity, $\psi=1$. Moving to the center-of-mass $R=(z_1+z_2)/2$ and relative $r=z_1-z_2$ coordinates, we can approximately relate $\psi(R,r)$ to the outgoing light by
\begin{equation}
\psi(R=L,r=v_\rmg\tau)=\sqrt{g^{(2)}(\tau)}{\rm e}^{{\rm i}\phi(\tau)}.\nonumber
\end{equation}
It can then be shown that the evolution of a stationary Poissonian input $\psi(R=0)=1$ in a dispersive nonlinear medium, assuming $\Omega_{\rm c}\ll\Dp$, is given approximately by a Schr\"odinger-like equation \cite{Vuletic2013b}
\begin{equation}
{\rm i}\frac{\partial \psi}{\partial R} =
\frac{4\la\Dp}{\gamma} \frac{\partial ^{2}\psi }{\partial r^{2}}+ \frac{\gamma}{\la|\Dp|} U(r)\psi.
\label{eq:RydSch}
\end{equation}
The center-of-mass coordinate, varying from $R=0$ to $R=L$, plays the role of time in this Schr\"odinger evolution. The first term on the right-hand side accounts for an effective photon mass. It  stems from the finite bandwidth of the linear EIT transmission, rendering a quadratic dispersion of the individual polaritons. The second term describes an effective potential
\begin{eqnarray}
&U(r)=\frac{1}{(r/\rb)^6 + {\rm sign}(\Dp)}=\left\{
\begin{array}{cc}
{\rm sign}(\Dp) & r\ll \rb \\
0 & r\gg \rb%
\end{array}%
\right.,\label{eq:effpot}
\end{eqnarray}
where ${\rm sign}(\Dp)=\Dp/|\Dp|$ and $\rb=\sqrt[6]{2C_6|\Dp|/(\hbar\Omega_{\rm c}^2)}$ (in an off-resonance EIT, where $\Geit=\Omega_{\rm c}^2/|4\Dp|$). Equation (\ref{eq:effpot}) assumes a repulsive van-der-Waals interaction between the Rydberg atoms, $C_6/r^6$ with $C_6>0$.

We thus find that the behaviour of a photon pair in a Rydberg-EIT medium resembles that of a pair of interacting massive particles. Since $\Dp$ determines the sign of both the mass and the potential, the overall behaviour is that of an \emph{attractive potential well}, or an attractive force, irrespective of the sign of $\Dp$. However, the well $U(r)$ is well-behaved only for $\Dp>0$; for $\Dp<0$, the denominator in Eq.~(\ref{eq:effpot}) vanishes at the boundaries ($|r|\approx \rb$) due to a resonant Raman absorption, creating local resonance-like features in $U(r)$.

For $\Dp>0$, bound-state solutions of Eq.~(\ref{eq:RydSch}) may be termed `molecules' of two photons. In experiments so far \cite{Vuletic2013b}, the well was shallow and supported only a single bound state $\psi_{\rm bound}$.
The "finite-time" evolution (from $R=0$ to $R=L$) following Eq.~(\ref{eq:RydSch}) is then governed by $\psi_{\rm bound}$. The initial state describing the lack of photon-photon correlations at the entrance to the medium is $\psi(R=0,r)=1$ ({\it i.e.}, a uniform distribution in the relative coordinate $r$), which is a superposition of a bound-state component $\psi_{\rm bound}$ and a scattering component $1-\psi_{\rm bound}$. The difference in the accumulated phase between these two components leads to constructive interference between them at $r=0$, and hence to photon bunching \cite{Vuletic2013b}.
This bunching and a reminiscence of the bound state are shown in Figs.~\ref{fig:dispersiveQNLO2}(a,c).

The case $\Dp<0$ was theoretically analyzed by Maghrebi, Gullans, {\it et al.}~\cite{Gorshkov2015b}. The resonance-like features in $U(r)$ resemble a Coulomb potential at $|r|\lesssim\rb$. They support a continuum of metastable bound states with an hydrogen-like energy spectrum, with the two polaritons separated by a finite distance $\sim\rb$. In the latter sense, this type of photonic 'molecule' perhaps resembles a real molecule more than the one at $\Dp>0$ (described above), which peaks at zero separation. However, the metastable states propagate in the medium with negative group velocity while decaying to a pair of Rydberg atoms at a rate $|\Omega_{\rm c}^2/\Dp|/\phi^2$ [where $\phi=-(\ODB/2)(\gamma/\Dp)$ as defined above].

\subsection{Correlated states: from two to many photons}
The successful realization of quantum nonlinearity with Rydberg-EIT prompted great theoretical efforts to better describe and understand the system. A full description of a uniform EIT system involves one dark and two bright polariton branches, obtained by diagonalizing the non-interacting Hamiltonian in momentum space. In this basis, the Rydberg interaction appears as a non-trivial local scatterer, coupling between the different branches. The scattering properties for the case of two photons in one dimension were derived by Bienias {\it et al.}~\cite{Buechler2014} using quantum scattering theory. This approach requires very little assumptions and is thus applicable for a wide range of parameters, including for large blockade radii and strong control fields $\Omega_{\rm c}$, on the order of or even much larger than $\Dp$. In particular, Bienias {\it et al.}~show that increasing $\Omega_{\rm c}$ can modify the 1D scattering length from attractive, as in Eq.~(\ref{eq:RydSch}), to repulsive. Such transition occurs at a scattering resonance similar to a Feshbach resonance. Furthermore, this approach provides a generalization of Eq.~(\ref{eq:RydSch}) to the non-stationary regime: For fields slowly-varying in time, one only needs to replace $\partial/\partial R\rightarrow\partial/\partial R+\partial/(v_\rmg\partial t)$, rendering a spatio-temporal Schr\"odinger-like dynamics. Finally, an effective \emph{many-body} Hamiltonian can be formulated in terms of the 1D scattering length for low energies.

An alternative approach to scattering theory is the input-output formalism. In this formalism, the system is described in the Heisenberg picture by defining a set of operators for the optical modes, including special operators for the input and output modes. For many quantum optics systems, this approach is more natural and proves extremely useful \cite{Collett1985,Fan2010}. Recently, Caneva {\it et al.} introduced a generalized input-output formalism for describing the Rydberg-EIT system \cite{Chang2015}. They effectively model the system as a one-dimensional chain of interacting three-level atoms (a spin model) that is tailored to reproduce the mean-field parameters of the real continuous medium, such as OD and $v_\rmg$. Calculating high-order correlations of the outgoing photonic state is then done in a relatively straight-forward procedure by solving for higher moments of the Heisenberg operators.

An exact many-body formulation of the continuous system was recently presented by Moos {\it et al.}~\cite{Fleischhauer2015}. The model includes photon loss from the bright polariton branches, arising from dark-bright coupling due to the Rydberg-Rydberg interaction. The effect of finite beam size (paraxial propagation) is also considered. Even so, an effective one-dimensional, many-body, model for only the dark polaritons is still valid under certain conditions, as verified by exact numerical simulations \cite{Fleischhauer2015}.

An exciting prospect for quantum nonlinear optical systems is their potential to realize strongly-correlated states and dynamical many-body phases with photons. While experiments are being setup to pursue this regime, there have been a few theoretical predictions for a many-body behavior.

Honer {\it et al.}~proposed a single-photon absorber based on an effective two level system (large $\Dp$), with the Rydberg state sufficiently interacting such that a single excitation blockades the complete optical medium \cite{Buechler2011}. In this case, after the first absorption event, the blockaded medium becomes completely transparent for all subsequent signal photons. If the optical depth is sufficient to absorb the first photon with large probability, this removes with high fidelity exactly one photon from an arbitrary input state. In the opposite limit of dissipative nonlinearity ($\Dp=0$), the system transmits single photons while scattering the multi-photon components. The back-action of this scattering on the properties of the transmitted photon was investigated by Gorshkov {\it et al.}~\cite{Pohl2013}. This work implies that when the medium is smaller than one blockade sphere, it will transform an intense coherent input into a stream of single photons with a well-defined separation.

Such a photonic state, a so-called one-dimensional crystal of photons, promises to be a valuable resource for metrology and quantum computation. In the absence of dissipation, with only dispersive (conservative) interaction between the photons, crystallization of an incoming coherent field is akin to Wigner crystallization of electrons. This many-body process, which requires dynamic control of the group-velocity, was studied by Otterbach {\it et al.}~using Luttinger liquid theory in the dilute, low-energy regime \cite{Fleischhauer2013} and afterwards validated by Moos {\it et al.}~using numerical simulations of their many-body model \cite{Fleischhauer2015}.

\section{Applications}
 \begin{figure}[t]
\begin{center}
\includegraphics[width=7 cm]{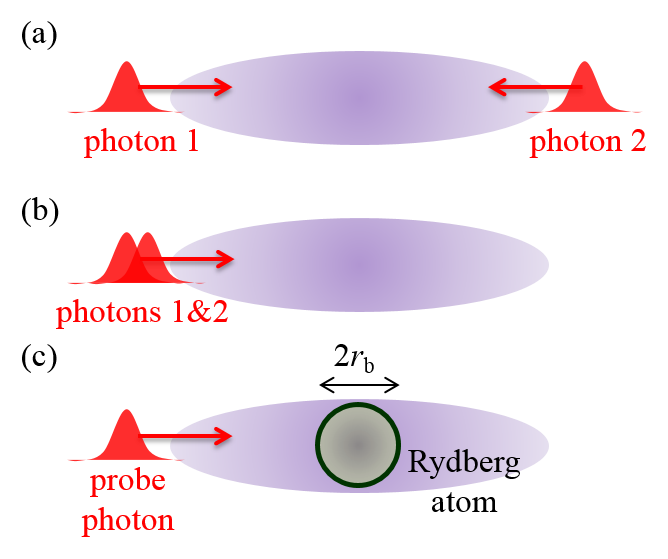}
\end{center}
\caption{Schematic of different schemes for making individual photons interact inside a Rydberg medium. Photons can interact while (a) counter-propagating or (b) co-propagating through the medium. (c) Alternatively, one (or more) photons can first be completely stopped and stored in the medium as a Rydberg atom and then interact with subsequent photons sent into the medium.}
\label{fig:photonphoton_schematic}
\end{figure}
The effective interaction between individual photons %due to the Rydberg-EIT mediated nonlinearity
discussed in the previous section enables a variety of optical quantum information applications. Since the first proposal %by Friedler {\it et al.}~in 2005
for a photonic phase gate using Rydberg-EIT \cite{Kurizki2005}, a range of different ideas have been suggested %proposed among others
for photonic quantum gates or non-classical light sources \cite{Kurizki2005,Lukin2011,Buechler2011}. The immense experimental progress in the last years has lead to a number of demonstrations of applications making use of the Rydberg interaction in an atomic medium.  The dissipative interaction (section \ref{subsec:dissipative_nonlinearities}) has been used to demonstrate highly efficient single-photon generation \cite{Kuzmich2012b,Adams2013}, atom-photon entanglement \cite{Kuzmich2013}, as well as single-photon all-optical switches \cite{Duerr2014} and transistors \cite{Hofferberth2014,Rempe2014b}. In turn, the dispersive nonlinearity (section \ref{subsec:dispersive_nonlinearities}) has been exploited to imprint large conditional phase shifts on weak target pulses. Very recently, a record phase-shift exceeding $\pi$ conditioned on the storage of a single gate photon has been reported by Tiarks {\it et al}. \cite{Duerr2016}.

All these application make use of the already discussed Rydberg EIT ladder scheme [Fig. \ref{fig:chi}(a)]. Single photons or weak coherent `target'
pulses on or near resonance with the $\vert {\rm g}\rangle\leftrightarrow\vert {\rm e}\rangle$ transition are sent into the optically thick medium, while a strong control field couples $\vert {\rm e}\rangle$ to the target Rydberg state $\vert {\rm r}\rangle$. For conditional operations, a second weak `gate' pulse is coupled either to the same or via a second control laser to a different Rydberg state $\vert {\rm r'}\rangle$ \cite{Duerr2014,Hofferberth2014,Rempe2014b}. Employing different Rydberg states simultaneously greatly enhances the flexibility of the implemented schemes as target and gate photons can for example be individually slowed and stored in the medium.
Various different schemes for achieving interaction either between different photons in the target pulse or between target and gate photons have been proposed and implemented (Fig. \ref{fig:photonphoton_schematic}). Friedler {\it et al}. initially considered two single photon pulses counter-propagating through the medium. For dispersive interaction this results in a phase-imprint during the "collision" of the two slowly-propagating polaritons, which maps into a phase-shift of the optical field outside the medium \cite{Kurizki2005}. Alternatively, signal and gate photons can co-propagate through the medium, in which case they will interact during their travel time through the medium \cite{Vuletic2012,Vuletic2013b}. While this can maximize the interaction time, it may be more challenging to separate signal and gate photons and thus to perform a controlled operation on a target pulse. Finally, the ability to completely stop a photon and convert it into a stored spin-wave inside the medium enables the configuration shown in Fig.~\ref{fig:photonphoton_schematic}(c): one photon is first stored and subsequently interacts with multiple photons propagating through the medium. This scheme is particularly suited for applications where a single gate photon should interact with many signal photons, as in the optical transistor applications \cite{Hofferberth2014,Rempe2014b}. One has to keep in mind though, that the retrieval of the stored gate photon, required for the realization of a full quantum gate, can be strongly affected by the interaction with the signal photons \cite{Lesanovsky2015,Hofferberth2015b}.

In the following, we review three specific applications, which are currently being explored both experimentally and theoretically, namely storage-based generation of non-classical light, photonic two-qubit phase gates, and all-optical switches and transistors.

\subsection{Generation of non-classical light from collective Rydberg excitations}
\label{subsec:generation_of_nonclassical_light}
\begin{figure}[tb]
\begin{center}
\includegraphics[width=0.7\columnwidth]{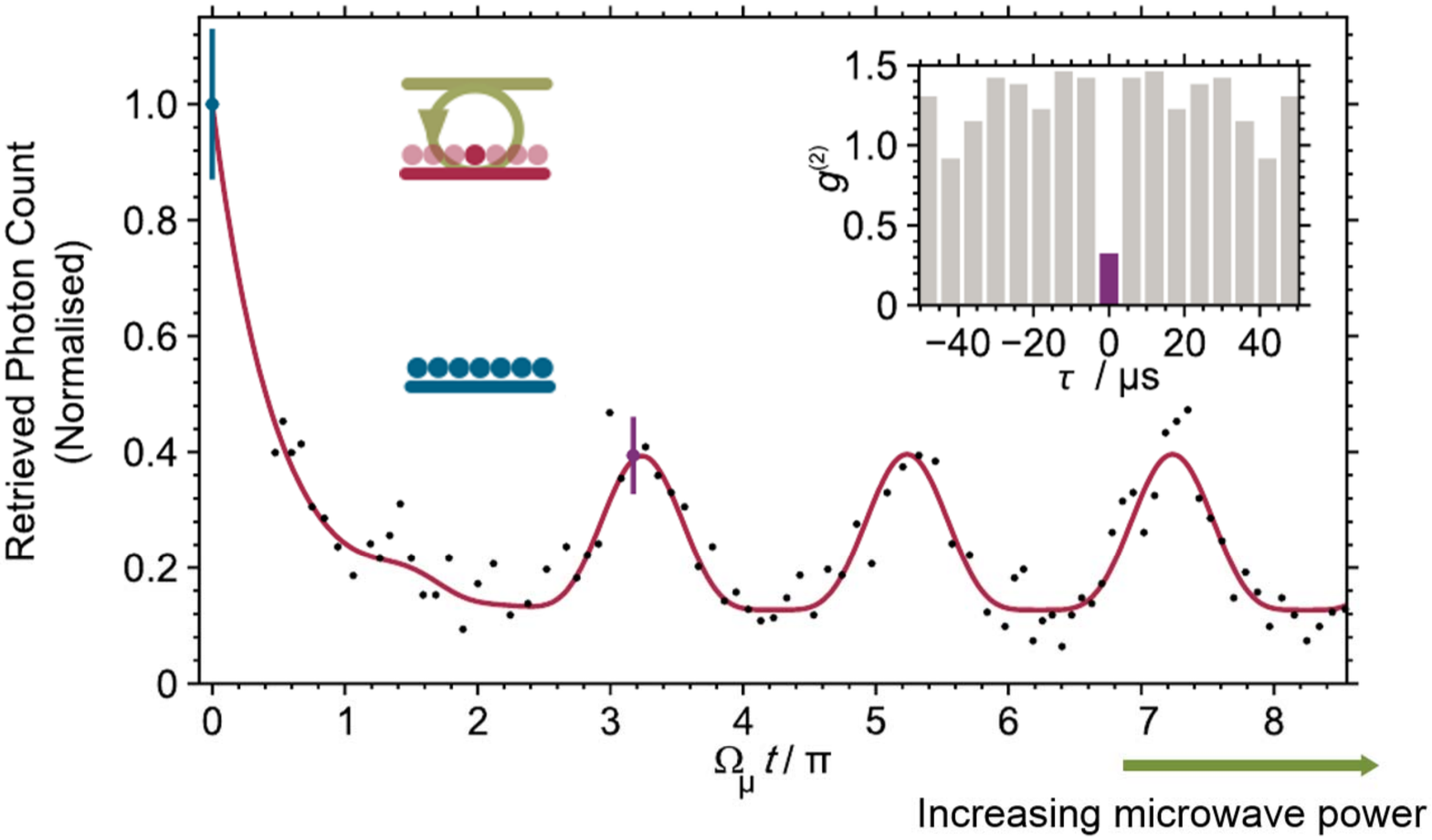}
\end{center}
\caption{Manipulation of light stored as a Rydberg excitation using an external microwave field \cite{Adams2013,Adams2014b}. The microwave field drives Rabi oscillations between the polariton Rydberg state and another Rydberg level as depicted schematically in the inset (upper left). This induces resonant dipole-dipole interactions which modify the photon statistics suppressing $g^{(2)}$ at the first revival in the retrieved pulse, inset (upper right). Figure adapted from Maxwell {\it et al.}~\cite{Adams2013,Adams2014b}.}
\label{fig:Adams_singlephoton_microwave}
\end{figure}
The concept of a Rydberg-mediated non-classical light source is to convert a weak coherent input pulse into non-classical output via the effective interaction inside the medium. In other words, the Rydberg medium acts as a filter for the input light, changing its photon statistics based on how the Rydberg interaction in the medium is coupled to the signal photons. The Rydberg blockade mechanism enables such operation with the aid of EIT and slow-light, but also outside the EIT regime.

The first quantum light source originating from a Rydberg excitation was demonstrated by Dudin and Kuzmich \cite{Kuzmich2012b} in 2012. A coherent signal pulse with large detuning $\Dp$ creates a collective Rydberg excitation inside a medium that is larger than a single blockade. The long-range interaction between different excitations results in strong dephasing of the initial many-atom state \cite{Kuzmich2012b,Grangier2012b}. As a consequence, when a resonant ($\Dp=0$) readout field is turned on after some time, the retrieved signal photons are scattered in random directions. On the other hand, if only a single excitation is created by the input beam, no dephasing occurs and a single photon is retrieved in the phase-matched direction. The single-photon character of the signal output was characterized in the experiment by measuring the second order correlation function. The reported value of $g^{(2)}(0) = 0.040(14)$ shows the extreme suppression of readout containing more than one photon. Furthermore, for an optical medium smaller than a single blockade volume, Dudin {\it et al.}~showed that this optical readout can be used to probe the dynamics of the collective Rydberg excitation \cite{Kuzmich2012c}.

This approach to filtering a coherent input field to achieve a non-classical output was further explored by Maxwell {\it et al}. \cite{Adams2013,Adams2014b}. In their experiment, the signal photons were stored as a collective excitations of a Rydberg $S$-state by turing off the control field during the probe pulse. The Rydberg excitation was subsequently coupled to a neighbouring $P$-state using a resonant microwave field before retrieving the stored light by turing the control field back on (Fig. \ref{fig:Adams_singlephoton_microwave}). Due to the microwave field, the interaction between excitations was tuned from van-der-Waals to dipolar, which changes both the strength and the angular dependence \cite{Cote2005,Walker2008,Hofferberth2015}. The resulting dynamics of the interacting Rydberg excitations \cite{Ates2013} became visible in the photon statistics of the retrieved signal light, as is shown in Fig. \ref{fig:Adams_singlephoton_microwave}. From this first demonstration it becomes apparent that the extreme tunability of the Rydberg interaction opens further possibilities for tailoring the stored light and for realizing more complex non-classical light states.

\subsection{Photonic phase gates}
\label{subsec:photonic_phase_gates}
To implement a two-photon phase gate, the dispersive interaction (section \ref{subsec:dispersive_nonlinearities}) between polaritons is used. The case of two counter-propagating polaritons was first discussed by Friedler {\it et al}. \cite{Kurizki2005} and further explored and extended to co-propagating and stored pulses by Gorshkov {\it et al.}~\cite{Lukin2011}. The basic idea is the same in all configurations: the interaction between two individual polaritons results in a $\pi$-phase shift imprinted on the transmitted light, while the large detuning from the intermediate state results in (ideally) zero scattering of the photons. To turn such a conditional phase shift into a quantum gate, the mechanism needs to be state-dependent with regard to whatever basis is used to encode the quantum information in individual photons. A straightforward example is the common polarization encoding, in which case selection rules can be used to couple only one specific combination of signal and gate photon polarization to the Rydberg state \cite{Vuletic2013b}.

A Rydberg-interaction mediated phase shift of an optical pulse was first reported by Parigi {\it et al.}~in 2012 employing an atomic ensemble inside an optical resonator \cite{Grangier2012}. Tiarks {\it et al.}~have very recently reported a conditional phase shift for single photons exceeding $\pi$, by making use of an optical medium with large optical density ${\rm OD}=25$ and storage and retrieval of the gate photon \cite{Duerr2016}. As explained in section \ref{subsec:dispersive_nonlinearities}, % unlike experiments only interested in photon number statistics,
 the characterization of the conditional phase-shift requires an interferometric measurement. Tiarks {\it et al.}~employ quantum state tomography in the polarization basis, similarly to that used by Firstenberg {\it et al.}~in the first demonstration of Rydberg-mediated dispersive interactions \cite{Vuletic2013b}.

Besides achieving record single-photon phase shifts, the Rydberg-mediated approach may overcome fundamental limitations of single-photon $\pi$ phase gates in conventional nonlinear media \cite{Shapiro2006,GeaBanacloche2010}. The key point here is that due to the long-range interaction the nonlinearity is no longer local, and the phase-shift can be uniform over the full size of the stored spin wave \cite{Adams2014}, which should allow a high-fidelity phase gate without the unavoidable pulse distortion for conventional nonlinearities.

\subsection{Optical switches and transistors}
\label{subsec:optical_switches}
\begin{figure}[t!]
\begin{center}
\includegraphics[width=0.7\columnwidth]{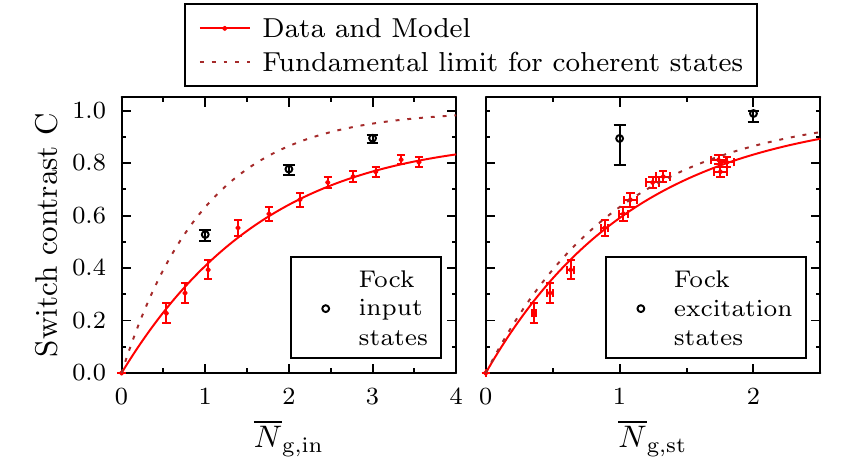}
\end{center}
\caption{(a) Switch contrast (red), i.e. the relative attenuation of the source beam, of a Rydberg mediated single photon transistor as function of mean photon number $\overline{N}_{g,in}$ in a coherent gate pulse. The experimentally observed attenuation of the source light is mainly determined by the Poissonian statistics of the gate photon number, with the dashed line giving the fundamental limit $C_{coh} = 1- e^{-\overline{N}_{g,in}}$ for a coherent input pulse. Knowing the gate photon statistics, the achievable switch contrast for one-, two- and three-photon Fock input states (black data points) can be predicted. (b) With the measured storage efficiency for individual gate photons, the data in (a) can be rescaled to show how much contrast is provided by gate excitations with Poissonian number distribution (red). This is again compared to the fundamental limit set by the gate excitation statistics. Block dots show the achievable contrast for deterministic single and two stored gate excitations. Figure adapted from Gorniaczyk {\it et al.}~\cite{Hofferberth2014}.}
\label{fig:Stuttgart_transistor}
\end{figure}
In a single photon switch the transmission of a target photon is controlled by a single gate photon. The first Rydberg-based switch was demonstrated by Baur {\it et al}. in 2014 \cite{Duerr2014}. The scheme is similar to the phase gate described in the previous section: the gate photon is converted into a stationary Rydberg excitation in the medium, either via storing of a Rydberg polariton or via direct excitation. %(large detuning $\Dp$ from the intermediate state).
The target photon is subsequently sent into the medium on resonant EIT coupled to a different Rydberg state. The gate excitation changes the optical response inside its blockade volume to that of an effective two-level system, see Eq.~(\ref{eq:t2}), resulting in scattering of the target photon from ground state atoms inside this volume. Since the blockaded optical depth $\ODB$ can be (much) larger than one, the Rydberg-based switch can achieve a very high on/off contrast $C>0.9$ for a single gate excitation (Fig. \ref{fig:Stuttgart_transistor}).

\begin{figure}[tb]
\begin{center}
\includegraphics[trim={5cm 8.5cm 5cm 8.5cm},clip,width=0.7\columnwidth]{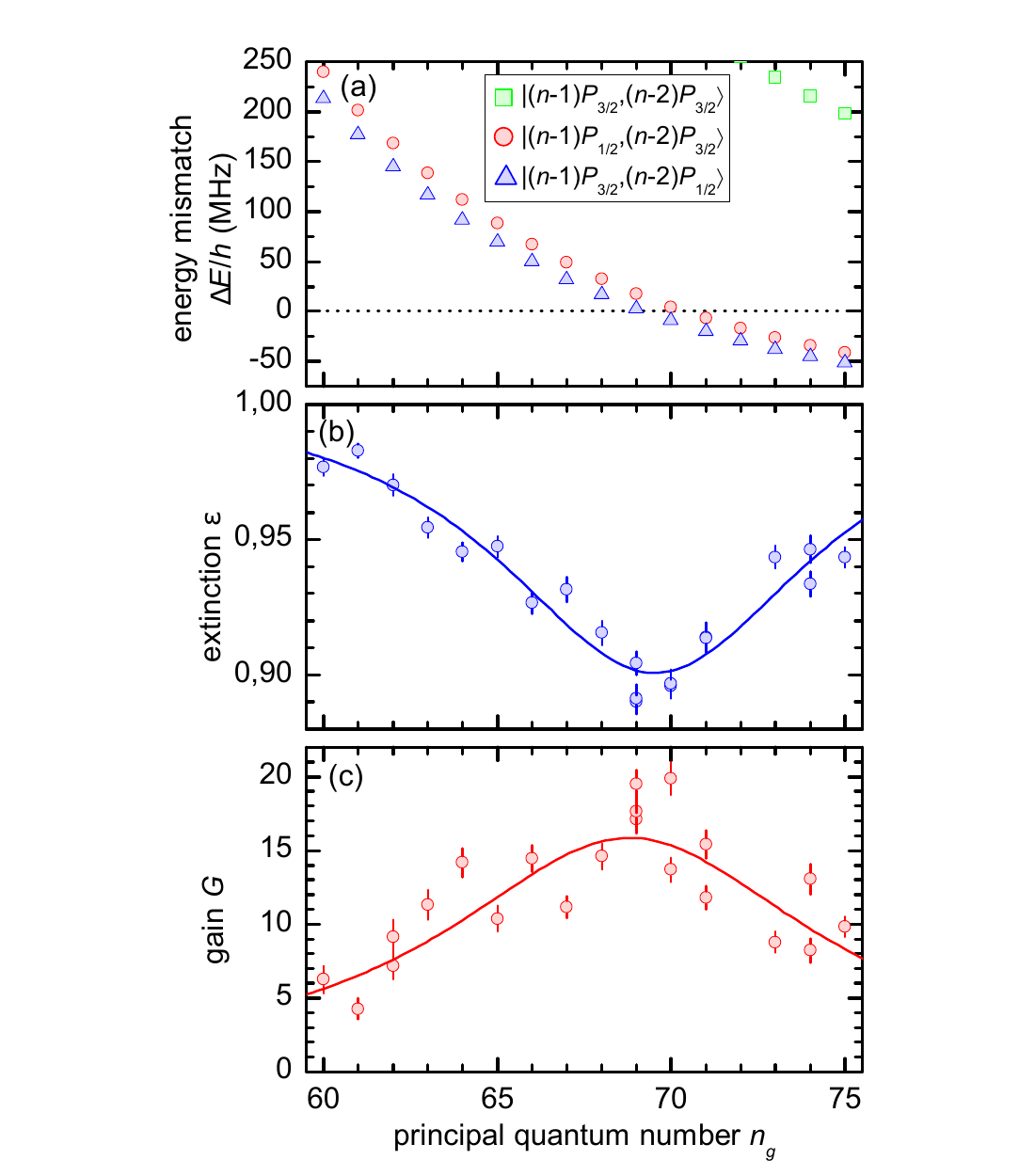}
\end{center}
\caption{The performance of the Rydberg single-photon transistor can be optimized solely by choosing the most appropriate Rydberg states to which gate and source photons are coupled. (a) Energy differnce ("F\"{o}rster defect") between the initial gate/source Rydberg state pair and the nearest dipole-coupled pair state. It can be seen that this defect becomes minimal for one specific choice of states. (b \& c) Measurements of the switch extinction and transistor gain as function of principal quantum number. The minimum in the F\"{o}rster defect results in maximal transistor performance. Figure adapted from Tiarks {\it et al.}~\cite{Rempe2014b}}.
\label{fig:Rempe_transistor}
\end{figure}

This concept was then used to demonstrate for the first time single photon transistors by Gorniaczyk {\it et al.}~\cite{Hofferberth2014} and Tiarks {\it et al.}~\cite{Rempe2014b}, where a single gate photon attenuates a stronger source input beam. This transistor performance is quantified by the optical gain $G$, which shows how many photons are removed from the source input by a single gate photon \cite{Vuletic2013}. Both experiments achieved a gain of $G \sim 15...20$.

Key to this achievement is the immense flexibility of the Rydberg interaction \cite{Cote2005,Shaffer2006,Walker2008}.  For the transistor experiments in particular, employing Rydberg states with different principal quantum numbers coupled to gate and source photons turned out to be an essential step. Practically, this avoids cross-talk between gate and source photons and excitations, but more importantly it enables tuning of the different interaction strengths involved in the scheme. For an optimal transistor, the interaction between the gate and source Rydberg state should be maximal, while the interaction among source photons is ideally small. One particular interesting feature for tuning the Rydberg interaction are F\"{o}rster resonances \cite{Martin2004,Pillet2006,Raithel2008,Entin2010,Pfau2012,Weidemueller2013b,Browaeys2015b}, which occur when two dipole-coupled pairs of Rydberg states are resonant with each other. This results in the transition from the van-der-Waals interaction regime $V_{\mathrm{vdW}} \sim C_6/r^6$ to dipolar interaction $V_{\mathrm{dd}} \sim C_3/r^3$. Tiarks {\it et al}. showed that the transistor performance improves when choosing gate and source Rydberg states which are close to such a resonance even in zero field (Fig. \ref{fig:Rempe_transistor}). Tuning the F\"{o}rster defect to exact zero by applying external fields has recently enabled further improvement of the transistor, demonstrating an optical gain $G \sim 200$ \cite{Hofferberth2015b}.

In these experiments, the transistor is operated classically, meaning that the gate photon is lost in the process. Similar to the electronic transistor, this classical device enables the amplification of signals, one example application being the high-fidelity all-optical detection of single Rydberg excitations \cite{Weidemueller2012,Lesanovsky2011}. With the optical gain now demonstrated, spatially resolved single-shot imaging of Rydberg excitations embedded in a background gas is feasible.
In contrast, retrieving the gate photon after the transistor operation constitutes the first step towards a quantum transistor. Such a quantum device with $G >2$ would enable quantum circuits with gain and feedback or creation of entangled multi-photon states. The finite coherence time of the stored gate spin-wave reduces the possible transistor operation time, but with the recent improvements, a gain $G>2$ could be demonstrated even if the gate photon is retrieved afterwards \cite{Hofferberth2015b}. At the moment, the fidelity of this coherent transistor is limited because the scattering of source photons results in projection of the stored gate spin wave \cite{Lesanovsky2015}. Similarly to the phase gates discussed in the last section, the long-range character of the interaction can in principle avoid this problem. For this, the blockade volume of the single gate excitation must exceed the total system size (and $\ODB \gg 1$), an experimentally challenging task.

\section{Challenges and outlook}
In less than a decade, Rydberg nonlinear optics has created new capabilities that were only dreamed of before. Previously, there were no optical media with a large nonlinearity at the single photon level and now there are. Ultra-cold Rydberg ensembles have been used to create bound states of photons, single photon phase shifts of order $\pi$, and all-optical transistors with gain larger than 100. These are remarkable successes but there remain considerable challenges both in terms of our theoretical understanding and practical applications.

\subsection{Current challenges and open questions}
A Rydberg EIT medium is a complex quantum many-body system where atoms couple collectively to a continuum of photonic modes. Light propagation, interactions and nonlinearity are all coupled. Currently, exact theoretical treatments are only able to describe interactions between two photons. Theoretical developments tend to advance hand in hand with experimental progress which provides a direct validation of approximations. With experiments starting to explore beyond pair-wise interaction, theory also must evolve from effective mean-field descriptions to true many-body treatment of the system \cite{Buechler2014,Fleischhauer2015,Chang2015}.

Much of the theoretical challenge in quantum nonlinear optics, as opposed, for example, to cold atoms and condensed-matter systems, stems from the optical nature of the system. It is naturally a nonequilibrium system, which is constantly driven and constantly monitored. It is also naturally dissipative, with the particle number (here, the photon number) not necessarily conserved. Therefore, to describe the evolution of highly-correlated states and nonequilibrium phases in a quantum nonlinear medium, modern methods of quantum field theory are required. These include scattering and input-output formalisms for quantised electromagnetic fields with multiple photons and multiple scatterers, dynamical many-body simulations, and other tools adopted from strongly-correlated condensed-matter theory. In turn, Rydberg-EIT systems of growing size could provide ideal testing grounds offering flexibility encountered in few other systems.

On the experimental side, probably the most fundamental challenge encountered so far is the limitation of the atomic density in which Rydberg excitations can be embedded without affecting their coherence or even lifetime \cite{Pfau2014c,Duerr2014}. Due to the large size of each Rydberg atom, at high atomic densities ($\gtrsim 10^{13}/$cc) surrounding ground state atoms are not only found inside the blockade volume, but even inside the orbit of the Rydberg electron. While this opens up access to highly interesting new physics such as low-energy electron-atom collisions and Rydberg molecule formation \cite{Sadeghpour2000,Pfau2009}, in the context of quantum nonlinear optics this imposes a major obstacle for increasing the strength of the effective photon-photon interaction. The increasing rate of Rydberg electron-atom collisions at higher atomic densities has been found to result in shorter coherence times \cite{Duerr2014} and even reduced Rydberg lifetimes \cite{Ott2015b}.

The key parameter encountered in virtually any application of Rydberg nonlinearities is $\ODB$. Many figures of merit depend strongly on $\ODB$ and only weakly on the total optical depth OD, such as the blockade probability for co-propagating photons $1-{\rm e}^{-\ODB}/\sqrt{\rm{OD}}$ \cite{Vuletic2012} or their conditional phase shift $\propto\sqrt{\rm{OD}}\cdot\ODB$ \cite{Vuletic2013b}. Other merits depend solely on $\ODB$, such as the attenuation of `signal' polaritons by a stored `gate' polariton in the photon transistor which equals $1-{\rm e}^{-\ODB}$ \cite{Hofferberth2014}. The parameters dominating the optical depth per blockade sphere $\ODB$ are the atomic density and the principle quantum number. Since the atomic density is fundamentally limited by the electron-atom collisions, the remaining knob for experimentalists is the principal quantum number. But this raises the same problem of electron-atom collisions, and there are practical issues such as available laser power and ability to compensate stray electric fields that have limited Rydberg-EIT experiments to principle quantum numbers $n \approx 100$. While this number will be pushed further upwards in future experiments, the practically achievable optical depth per blockade region will not grow much beyond $\ODB \sim 20$.

\subsection{Future directions}
To circumvent the limitations on $\ODB$, various attempts are currently in progress to implement more intricate schemes: Including, for example, the idea of enhancing the optical cross section using magic monolayers \cite{Adams2016}. Another option to further increase the photon-photon interaction is the introduction of an optical cavity around the atomic ensemble. A ground-breaking experiment by Parigi {\it et al}. provided a proof of concept for such a system in 2012 \cite{Grangier2012}, recently followed up by the demonstration of long-lived cavity-Rydberg polaritons by Ningyuan {\it et al}. \cite{Simon2015} and detailed study of Rydberg-induced nonlinearities inside a resonator by Boddeda {\it et al}. \cite{Grangier2016}. In parallel, theoretical analyses of this system suggest promising predictions for controlling the quantum statistics of light \cite{Grangier2014b,Grangier2015,Shaffer2016}, high-fidelity conditional-phase gates \cite{Sorensen2016}, and the realization of quantum crystals and fractional quantum Hall states of light \cite{Gorshkov2015,Simon2015b}. The latter make use of the rich 2D transverse-mode spectra of multimode cavities. While the introduction of an optical resonator makes experimental setups somewhat more complicated again, the new physics offered by these combined systems easily justifies this addition. It seems more than likely that fundamental new steps will emerge from these systems.

Another exciting research direction is to transfer the concepts discussed in this review and explored in ultra-cold atomic samples to room-temperature vapour cells. This could result in a great simplification of the experimental setup and pave the way for employing Rydberg-based nonlinearities in future devices. Rydberg interaction effects have already been observed in room-temperature experiments \cite{Carr2012,Pfau2013b,Loew2015}, suggesting that the concepts of quantum nonlinear optics with Rydberg atoms can be transferred from the ultra-cold. So far, experimental demonstration of light manipulation on the quantum level is still outstanding, but it seems safe to expect further progress in the future.

Based on these open challenges and future steps, the field of Rydberg-based quantum nonlinear optics will keep both theorists and experimentalists busy for quite some time. Certainly we can expect both fundamental breakthroughs, as well as paradigm shifting applications in photonic quantum information processing in the upcoming years.

\emph{Acknowledgments}.--- We are particularly grateful to Hans Peter B\"{u}chler, Stephan D\"{u}rr, Michael Fleischhauer, Tom Gallagher, Alexey Gorshkov, Philippe Grangier, Igor Lesanovsky, Misha Lukin, Klaus Molmer, Tilman Pfau, Pierre Pillet, Thomas Pohl, Gerhard Rempe, Vladan Vuletic, and many others for their insightful contributions that helped to shape this field over the last decade. O. F. acknowledges financial support from the Israel Science Foundation, the Laboratory in Memory of Leon and Blacky Broder, the Sir Charles Clore research prize and the MINERVA Stiftung with the funds from the BMBF of the Federal Republic of Germany. C. S. A. acknowledges financial support from EPSRC Grant Ref. No. EP/M014398/1, the EU RIA project `RYSQ' project and DSTL. S. H. is supported by the German Research Foundation through Emmy-Noether-grant HO 4787/1-1 and within the SFB/TRR21, and by RiSC grant 33-7533.-30-10/37/1 from the Ministry of Science, Research and the Arts of Baden-Württemberg.

\section{References}
\bibliographystyle{iopart-num}
\bibliography{biblio_EITReview}
\end{document}